\documentclass[10pt,conference]{IEEEtran}
\IEEEoverridecommandlockouts

\usepackage{url} 
\usepackage{framed}
\usepackage{mdframed}
\usepackage{listings}
\usepackage{multicol}
\usepackage{amssymb}
\usepackage{lipsum}
\usepackage{adjustbox}

\usepackage[font=bf,skip=4pt]{caption}
\usepackage{tabularx}
\usepackage{seqsplit}
\usepackage{multirow}
\usepackage{booktabs}
\usepackage{graphicx}
\usepackage{hyperref}
\usepackage{subcaption}
\usepackage{makecell}
\usepackage{xcolor}
\usepackage[ruled,vlined,linesnumbered] {algorithm2e}
\usepackage{amsmath}
\usepackage{tikz}
\usepackage{tcolorbox}
\tcbuselibrary{skins,breakable}
\usepackage{threeparttable}
\usepackage{tablefootnote}
\usepackage{csquotes}
\usepackage{colortbl}
\usepackage{algorithm2e}
\usepackage{adjustbox}
\usepackage{balance}
\usepackage{dblfloatfix}
\usepackage{placeins}
\usepackage{afterpage}

\usepackage{breakurl} 
\usepackage{boldline}
\usepackage[ruled,vlined]{algorithm2e}
 
\usepackage{pifont}%
\usepackage{balance}

\usepackage{enumitem}
\usepackage[htt]{hyphenat}

\newcommand{\find}[1]{
\begin{tcolorbox}[leftrule=0.5mm,toprule=0mm,bottomrule=0mm,
  left=0.7pt,right=0.7pt,top=0.2pt,bottom=0.2pt,
  colback=gray!10]
\em #1
\end{tcolorbox}
}

\newcommand{\findblue}[1]{
\begin{tcolorbox}[leftrule=0.5mm,toprule=0mm,bottomrule=0mm,rightrule=0.5mm,
  left=0.7pt,right=0.7pt,top=0.2pt,bottom=0.2pt,
  colframe=blue!60, colback=blue!8]
\em #1
\end{tcolorbox}
}

\newcommand{\findred}[1]{
\begin{tcolorbox}[leftrule=0.5mm,toprule=0mm,bottomrule=0mm,rightrule=0.5mm,
  left=0.7pt,right=0.7pt,top=0.2pt,bottom=0.2pt,
  colframe=red!80, colback=red!8]
\em #1
\end{tcolorbox}
}

\newcommand{\findorange}[1]{
\begin{tcolorbox}[leftrule=0.5mm,toprule=0mm,bottomrule=0mm,rightrule=0.5mm,
  left=0.7pt,right=0.7pt,top=0.2pt,bottom=0.2pt,
  colframe=brown!80, colback=brown!8]
\em #1
\end{tcolorbox}
}

\definecolor{customcite}{HTML}{b83b5e}
\definecolor{customlink}{HTML}{07689f}
\definecolor{customurl}{HTML}{b83b5e}
\AtEndPreamble{
\usepackage{hyperref}

}

\definecolor{customcell}{HTML}{ffebb7}

\newcommand{\tool}{\textsc{DagVul}}

\begin{document}

\date{}
\title{Beyond True or False: Why You Can't Trust the Reasoning in LLM-Generated Vulnerability Reports}

\author{
\IEEEauthorblockN{
Li Lu\textsuperscript{1},
Yanjie Zhao\textsuperscript{1}\ding{41},
Hongzhou Rao\textsuperscript{1},
Kechi Zhang\textsuperscript{2},
Zhou Yang\textsuperscript{3},
and Haoyu Wang\textsuperscript{1}
}
\IEEEauthorblockA{\textsuperscript{1}Huazhong University of Science and Technology, Wuhan, China\\
\{li\_lu, yanjie\_zhao, rhz, haoyuwang\}@hust.edu.cn}
\IEEEauthorblockA{\textsuperscript{2}Peking University, Beijing, China, zhangkechi@pku.edu.cn}
\IEEEauthorblockA{\textsuperscript{3}University of Alberta, Alberta Machine Intelligence Institute, 
Canada CIFAR AI Chair, Alberta, Canada, zy25@ualberta.ca}

\thanks{\ding{41} Yanjie Zhao is the corresponding author (yanjie\_zhao@hust.edu.cn).}

\thanks{\textsuperscript{1} The full name of the author's affiliation is Hubei Key Laboratory of Distributed System Security, Hubei Engineering Research Center on Big Data Security, School of Cyber Science and Engineering, Huazhong University of Science and Technology.}

}

\maketitle

\begin{abstract}

When a state-of-the-art LLM labels a function as vulnerable, can we trust its explanation?
This paper first reveals a worrisome fact: 30.3\% of correct verdicts from LLMs are produced through \textit{hallucinated or causally wrong reasoning}.
While prior work has partially observed this phenomenon, no study has quantified its prevalence, taxonomized its failure modes, or proposed a structural fix.

We aim to address all three. 
First, we construct reasoning-level datasets: real-world vulnerabilities across 149 CWE types annotated with evidence-based reasoning chains, and semantically equivalent code perturbations for robustness testing.
Through open coding of 400 reasoning traces across 10 LLMs, we identify 12 systematic failure modes, among which code comprehension and logical reasoning errors are the most prevalent.
Motivated by these findings, we propose \tool{}, which models vulnerability reasoning as a Directed Acyclic Graph (DAG) generation task, enforcing causal dependencies between reasoning steps and aligning them with program semantics via Reinforcement Learning with Verifiable Rewards (RLVR).
\tool{} improves reasoning F1-score by an average of 13.3\% over baselines. Our 8B-parameter model achieves performance comparable to Claude-Sonnet-4.5 under CWE-Guided prompting (78.90\% vs. 76.11\%) in the evaluated setting. These results support the combined benefits of structured causal reasoning and training on curated, task-specific data for vulnerability analysis.

\end{abstract}


\section{Introduction}
\label{sec:introduction}

Modern software security practices do not operate on binary signals alone. When a security engineer receives a vulnerability report, what matters is not merely the verdict \textit{vulnerable} or \textit{not vulnerable}, but the reasoning: which code path is exploited, what the root cause is, and how an attacker could trigger the flaw. 
Real-world vulnerability management~\cite{Zhou2025Large} unfolds as a staged pipeline: (Stage 1-detection) surface candidate flaws; (Stage 2-analysis) construct a reasoning chain explaining \textit{why} and \textit{how} a flaw is exploitable, typically in the form of a human-readable report; and (Stage 3-verification) validate that understanding through proof-of-concept (PoC) exploits or patches. This paper focuses on the first two stages, which are essential for Stage 3: a PoC exploit or patch cannot be valid if it is built on a misidentified root cause~\cite{Vul-R2}.

Large Language Models (LLM)-based tools are increasingly used to automate these stages~\cite{xu2025survey}.
LLMs can identify complex semantic dependencies and discover zero-day vulnerabilities that traditional SAST and DAST tools fail to detect~\cite{li2023hitchhikersguideprogramanalysis, Chen2025When, Li2024Enhancing}, achieving state-of-the-art performance in determining whether code is vulnerable~\cite{R2Vul}. 
Yet a correct verdict does not imply correct understanding: 
LLMs frequently produce \textit{correct binary verdicts} based on \textit{wrong reasoning}, misidentifying the true root cause while arriving at the correct label~\cite{SecLLMHolmes}. Practitioners acting on such flawed, LLM-generated reports may chase phantom fixes and overlook the real attack surface, leading to potentially catastrophic consequences.

Why has this problem persisted? A key reason is that vulnerability detection has largely been treated as an outcome-only task. Benchmarks tend to evaluate binary verdicts rather than reasoning quality. Training objectives typically reward correct labels rather than correct analysis. And the limited availability of expert-verified reasoning ground truth makes it difficult both to measure the problem and to address it~\cite{DBLP:journals/corr/abs-2506-05692,DBLP:conf/naacl/WangLX25a}. These factors reinforce each other: without reasoning-level evaluation, flawed reasoning is less likely to be noticed, which in turn reduces the incentive to collect reasoning data or rethink training objectives.

To bridge these gaps, we first construct two comprehensive datasets needed for reasoning-level evaluation. The first contains real-world vulnerabilities across 149 CWE types, each annotated with evidence-based causal reasoning chains as ground truth. The second applies semantically equivalent code transformations to test whether a model's reasoning is grounded in actual vulnerability logic or merely in surface-level code features.
Our empirical study uncovers a widespread \textit{``Flawed Reasoning, Correct Answer''} phenomenon: 30.3\% of correct binary verdicts are grounded in wrong logic rather than sound causal reasoning. 
Through open coding of 400 reasoning traces, we systematically taxonomize 12 distinct reasoning failure modes, with code comprehension and logical reasoning errors as the dominant categories.

These failures reveal a key challenge in current LLM vulnerability analysis: when a model reasons step-by-step in sequence (i.e., linear chain-of-thought (CoT) reasoning), it struggles to maintain evidence-grounded consistency across multi-step vulnerability reasoning under complex code semantics.
Thus, we propose \tool{}, a novel framework that reframes vulnerability reasoning as a Directed Acyclic Graph (DAG) generation task, where each reasoning step must explicitly declare its dependencies on prior steps.
This structure prevents unsupported inferential leaps and ensures that every conclusion is traceable to established code facts.
To address the outcome-only training objective identified earlier, we train the model with a reward signal that verifies both the structural completeness of the reasoning chain and the correctness of the final conclusion, rather than rewarding correct labels alone.

Evaluations show that \tool{} improves reasoning F1-score by an average of 13.3\% over strong baselines from recent works, including RAG~\cite{Fan2024RAG} (following Vul-RAG~\cite{Vul-RAG}), SFT~\cite{rajani2019explain} (following VulnLLM-R~\cite{VulnLLM-R}), and ORPO~\cite{hong2024orpo} (following R2Vul~\cite{R2Vul}). 
By enforcing strict evidence-grounded reasoning, \tool{} significantly reduces false alarms by an average of 24.79\%. 
Notably, our 8B model surpasses specialized large-scale reasoning models such as Qwen3-30B-Reasoning, and remains competitive with closed-source models such as Claude-Sonnet-4.5~\cite{anthropic2025sonnet45}.

\begin{figure}[!t]
    \centering
    \begin{tcolorbox}[
        colback=white, 
        colframe=black!70, 
        boxrule=0.8pt,
        arc=2mm, 
        width=\linewidth, 
        sidebyside, 
        sidebyside align=top, 
        sidebyside gap=2mm,
        lefthand ratio=0.52, 
        lower separated=true
    ]
    
    {\fontsize{5.8}{7}\selectfont 
    \textbf{A. Real-World Source Code} \vspace{1.5mm}\\
    \texttt{static void get\_conn\_text(...) \{} \\
    \texttt{~~char addr\_text[MAXPATHLEN];} \\
    \texttt{~~...} \\
    \texttt{~~case AF\_UNIX:} \\
    \textcolor{red!80!black}{\texttt{// True Bug: OOB Read}} \\
    \texttt{~~~~\colorbox{red!20}{strncpy}(addr\_text, ...);} \\
    \texttt{~~...} \\
    \textcolor{blue!80!black}{\texttt{// Wrong target}} \\
    \texttt{~~\colorbox{blue!10}{sprintf}(addr, ...);} \\
    \texttt{\}}
    }
    \tcblower 
    
    {\fontsize{6.5}{8}\selectfont
    \textbf{B. Claude-Sonnet-4.5's Output} \vspace{1.5mm}\\
    \textbf{VERDICT:} \textcolor{green!60!black}{yes $\checkmark$} \\
    \textbf{REASON:} \textcolor{red!80!black}{Flawed $\times$} \vspace{1.5mm}\\
    \textbf{1. Wrong Root Cause:} \\
    The final \colorbox{blue!10}{\texttt{sprintf}} writes without bounds checking... \vspace{1.5mm}\\
    \textbf{2. Fatal Misjudgment:} \\
    \colorbox{red!20}{\texttt{strncpy}} usage partially mitigates overflow... which is \colorbox{yellow!50}{\textbf{good practice}} here.
    }
    \end{tcolorbox}
    \caption{Motivation Example. The LLM correctly flags the function as vulnerable but attributes it to \texttt{sprintf}, while praising the vulnerable \texttt{strncpy} as ``good practice.''}
    \label{fig:motivation_example}
\end{figure}

In summary, our contributions are as follows:

\begin{itemize}[leftmargin=*]

    \item \textbf{Datasets.} We construct comprehensive reasoning-level datasets for LLM-based vulnerability detection: 4,998 CVEs across 149 CWE types annotated with evidence-based causal reasoning chains, paired with 100 manually verified semantic-preserving code transformations that isolate reasoning from superficial patterns.

    \item \textbf{Empirical Evidence.} We evaluate 10 LLMs and reveal that 30.3\% of correct verdicts are produced through flawed reasoning. We taxonomize 12 failure modes through open coding of 400 reasoning traces, showing that the main challenges lie in code comprehension and logical reasoning.

    \item \textbf{Technical Contribution.} We propose \tool{}, which models vulnerability reasoning as DAG generation with explicit step dependencies, and trains the model using rewards for logical closeness and reasoning correctness.
    Our 8B-parameter \tool{} outperforms larger open-source reasoning models and remains competitive with Claude-Sonnet-4.5 under CWE-Guided prompting (78.90\% vs. 76.11\%).
    
\end{itemize}
\begin{figure*}[t!]
    \centering
    \includegraphics[width=0.8\linewidth]{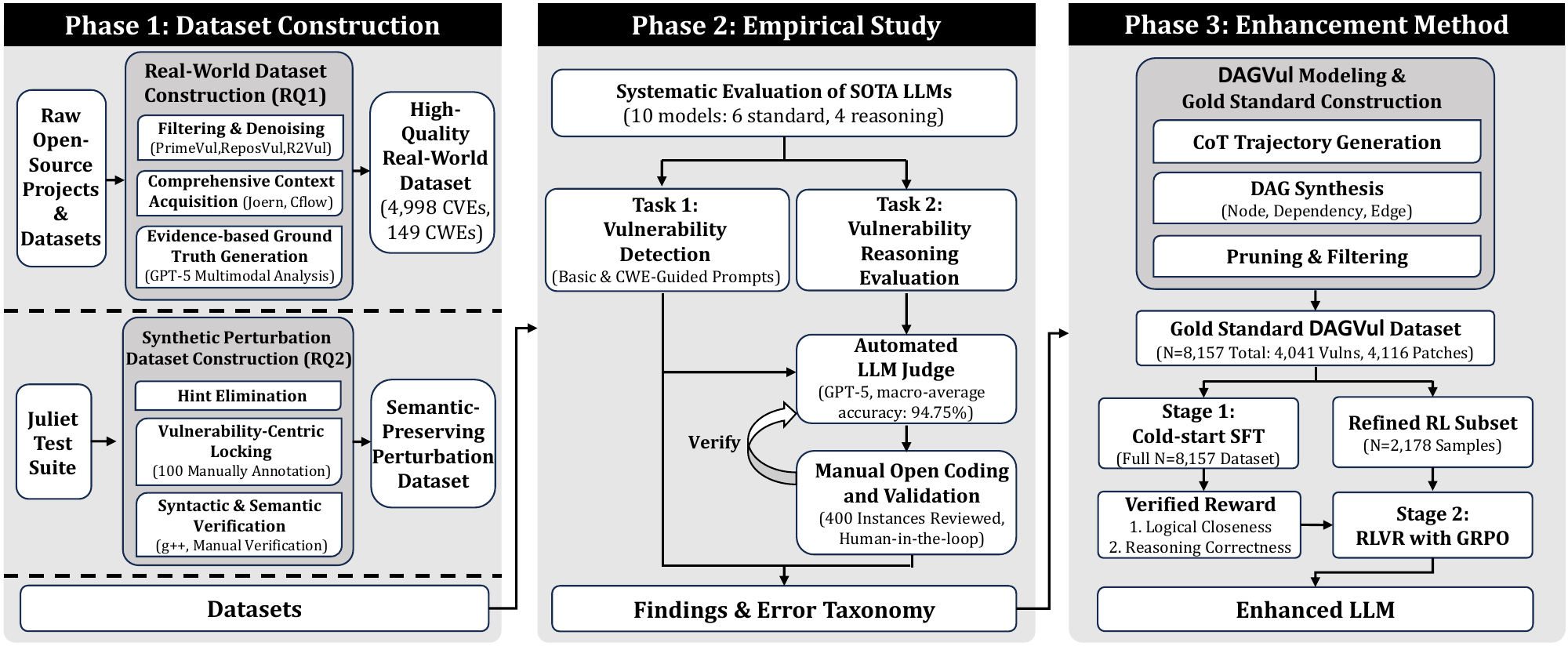}
    \caption{Overview of our framework.}
    \label{fig:overview}
\end{figure*}

\definecolor{cmarkcolor}{HTML}{46C45E} 
\definecolor{xmarkcolor}{HTML}{C0504D} 
\definecolor{hmarkcolor}{HTML}{D3A625} 

\newcommand{\cmark}{\textcolor{cmarkcolor}{\ding{51}}}
\newcommand{\xmark}{\textcolor{xmarkcolor}{\ding{55}}}
\newcommand{\hmark}{\textcolor{hmarkcolor}{\ding{108}}} 

\begin{table}[!t]
    \centering
    \caption{Recent evaluations for vulnerability detection. (\cmark = Supported, \xmark = Not Supported, \hmark = Partially Supported).}
    
    \resizebox{\linewidth}{!}{
        \begin{tabular}{l c c c c c c}
            \toprule[1.2pt]

            \textbf{Study} & 
            \makecell{\textbf{Real-}\\\textbf{World}\\\textbf{Dataset}} & 
            \makecell{\textbf{Label}\\\textbf{Accuracy}} &    
            \makecell{\textbf{Comprehensive}\\\textbf{Context}} &                 
            \makecell{\textbf{Evidence-based}\\\textbf{Ground}\\\textbf{Truth}} &  
            \makecell{\textbf{Perturb}\\\textbf{Method}} &              
            \makecell{\textbf{Open}\\\textbf{Coding}} \\                
            \midrule

            SecLLMHolmes~\cite{SecLLMHolmes}   & \xmark & \cmark & \xmark & \hmark & \cmark & \xmark \\
            CORRECT~\cite{CORRECT}        & \cmark & \cmark & \cmark & \hmark & \xmark & \xmark \\
            R2Vul~\cite{R2Vul}          & \cmark & \cmark & \xmark & \xmark & \xmark & \hmark \\
            SECVULEVAL~\cite{SECVULEVAL}     & \cmark & \cmark & \cmark & \xmark & \xmark & \hmark \\
            ReVD~\cite{ReVD}           & \cmark & \hmark & \xmark & \xmark & \xmark & \xmark \\
            SV-TrustEval-C~\cite{SV-TrustEval-C} & \xmark & \cmark & \xmark & \hmark & \cmark & \xmark \\
            VulnLLM-R~\cite{VulnLLM-R}      & \cmark & \cmark & \hmark & \xmark & \xmark & \xmark \\

            \bottomrule[1.2pt]
        \end{tabular}
    }
    \label{tab:benchmark_comparison}
\end{table}

\section{Background and Related Work}
\label{sec:formulation}

The default way to evaluate a security analysis tool is to ask: \textit{does it find bugs?} In recent LLM evaluations, this has been the dominant practice: report a verdict, measure accuracy, publish a number. Yet this framing conceals a deeper, more troubling question: \textit{does the tool find the bug for the right reasons?}

\autoref{fig:motivation_example} shows a vulnerability report generated by Claude-Sonnet-4.5. The actual vulnerability is an out-of-bounds read caused by \texttt{strncpy} in the \texttt{AF\_UNIX} branch. The model correctly labels the function as vulnerable, but its reasoning is wrong: it identifies \texttt{sprintf} as the main issue and describes the vulnerable \texttt{strncpy} usage as a ``partial mitigation'' and ``good practice.'' 
A developer following this explanation would likely fix \texttt{sprintf} and leave the real bug untouched. 
This gap is what our work sets out to address.

\subsection{Task Definition}
\label{sec:taskdef}
Automated vulnerability analysis is modeled as a binary classification problem. Given a code snippet $C$, the goal is to learn a mapping $f: C \rightarrow Y$, where $Y \in \{0, 1\}$ indicates the absence or presence of vulnerabilities. Existing benchmarks~\cite{BigVUl,CVEfixes,Chen2023DiverseVul,PrimeVul} primarily evaluate models by maximizing $P(Y \mid C)$ and reporting the F1 score.

We argue this paradigm is \textit{fundamentally incomplete.} 
Developers do not lack bug reports with merely a binary verdict.
What they lack are \textbf{actionable} ones with meaningful diagnoses~\cite{8530713}.
Moreover, a model that arrives at the correct label through surface-pattern matching rather than genuine code comprehension is inherently fragile. 
Semantic-preserving transformations can significantly degrade the model performance~\cite{alert}.
We therefore refine the task to \textit{vulnerability reasoning}: the model must generate a tuple $(Y, R)$, where $Y$ is the binary classification verdict and $R$ is an explicit reasoning trace that identifies the root cause and the triggering logic. The optimization objective thus shifts from maximizing $P(Y \mid C)$ to maximizing the joint probability $P(Y, R \mid C)$.

Thus, \textbf{reasoning success} requires satisfying two conditions: the verdict must be correct ($\hat{Y} = Y$), \textit{and} the reasoning trace must align with expert ground truth ($\text{Verify}(\hat{R}, R_{gt}) = \texttt{True}$). A correct verdict backed by misaligned or incorrect reasoning is treated as a failure.

\subsection{Review of Existing Evaluations}

Recent years have seen encouraging progress toward richer vulnerability evaluation, summarized in~\autoref{tab:benchmark_comparison}.
However, several gaps persist, particularly in reasoning assessment.

First, regarding \textbf{comprehensive context}, while recent datasets~\cite{R2Vul,ReVD} have significantly improved label accuracy via techniques like patch untangling, they are largely confined to isolated functions or files. This highlights the need to incorporate the repo-level dependencies necessary for evaluating reasoning.
Second, concerning \textbf{evidence-based ground truth}, although some works~\cite{SecLLMHolmes,CORRECT} attempt to provide textual rationales, they often rely on general CVE descriptions or broad CWE categories. Evidence-based root-cause reasoning is still required to serve as ground truth for assessing reasoning quality.
Third, in terms of \textbf{reasoning robustness}, few works~\cite{SecLLMHolmes,SV-TrustEval-C} incorporate systematic semantic-preserving code perturbations. Such perturbations are essential to reliably distinguish a model's genuine code comprehension from superficial pattern matching.
Finally, there remains a gap in \textbf{qualitative taxonomy}. Most works~\cite{SecLLMHolmes,CORRECT,ReVD,SV-TrustEval-C,VulnLLM-R} focus solely on quantitative metrics but rarely employ systematic qualitative methodologies, such as open coding~\cite{khandkar2009open,Wang2025Towards}, to deeply analyze \textit{how} and \textit{why} reasoning failures arise.

\findorange{\textbf{Key Observation: A Measurement Gap, Not Just a Performance Gap.} Existing benchmarks measure whether LLMs \emph{guess} correctly. Our work asks whether they \emph{reason} correctly, a fundamentally higher bar that current infrastructure cannot evaluate.
}

\section{Dataset Construction}
\label{sec:benchmark}

This section describes the dataset we constructed specifically for evaluating the task defined in \autoref{sec:taskdef}.
The dataset is the foundation in our overall study design~\autoref{fig:overview}.

\subsection{Data Filtering}
We aim to build a large-scale, high-quality vulnerability dataset with accurate labels and sufficient context. We draw from three recent sources that address known data quality issues in earlier datasets~\cite{BigVUl,CVEfixes,Chen2023DiverseVul} such as label noise and tangled patches~\cite{li2025cleanvulautomaticfunctionlevelvulnerability}: PrimeVul~\cite{PrimeVul}, ReposVul~\cite{ReposVul}, and R2Vul~\cite{R2Vul}.
These datasets employ rigorous vulnerability untangling strategies, including: 
(1) \textit{Single-function principle}, retaining only commits that modify a single function (used specifically by PrimeVul, while the others include cross-function vulnerabilities); 
(2) \textit{NVD-based verification}, cross-referencing function names and file paths with official CVE descriptions; 
(3) \textit{LLM-assisted filtration}, utilizing LLMs for relevance scoring followed by human sampling validation; and 
(4) \textit{Static analysis filtering}, leveraging tools (e.g., Infer~\cite{FacebookInfer}) combined with commit tracing to verify vulnerability presence.

We merge these three datasets and apply three additional filtering steps:
(1) \textit{Scope filtering} excludes non-source files (e.g., documentation, configurations) and test files.
(2) \textit{Length constraint} discard samples exceeding a context window of 16,384 tokens to fit mainstream LLMs;
(3) \textit{De-duplication} leverages established practices~\cite{SECVULEVAL,SecLLMHolmes} to remove identical samples via MD5 hashing on normalized code.
The resulting dataset comprises 5,078 CVEs across 149 CWE types.

\subsection{Interprocedural Context Acquisition}
\label{sec:3.2}
The filtered samples provide the vulnerable functions but suffer from incomplete context, typically containing only function calls. 
To further capture the interprocedural dependencies that drive complex bugs, we first clone the complete source code of all 805 associated repositories locally via their attached GitHub links. 
Then we build a pipeline on top of works from Joern~\cite{joern2024} and Cflow~\cite{gnucflow2021}.
The pipeline performs three tasks.
(1) \textit{Bidirectional call graph analysis} extracts direct callers with a recursion depth of three, revealing the full multi-layer impact scope beyond the simple callee-only approaches of earlier studies~\cite{ReposVul}.
(2) \textit{Recursive type resolution} traces \texttt{typedef} aliases to their original \texttt{struct}/\texttt{union}/\texttt{enum} definitions, resolving the incomplete-type problem endemic in prior static analysis pipelines~\cite{CORRECT}.
(3) \textit{Global state tracking} captures globally accessible variables modified or accessed by relevant functions, providing a comprehensive view of shared mutable state.
The pipeline assembles \texttt{calleeMethods}, \texttt{callerMethods}, \texttt{globalVars}, \texttt{importLibs}, \texttt{typeDefs}, and pre-/post-patch 
\texttt{vulnerableMethods} into context.

\subsection{Evidence-based Ground Truth Construction}
\label{sec:3.3}
With clean samples and their interprocedural context, we now construct reasoning-level ground truth for each vulnerability. Current benchmarks predominantly use NVD descriptions as ground truth~\cite{Le2022Survey,CORRECT}, but these are too brief to serve as a reference for evaluating reasoning. Since our source datasets map each CVE to its GitHub fix commit, we can trace each sample back to its associated pull request and linked issues, gaining access to developer discussions, reviewer comments, and in some cases, visual evidence (e.g., screenshots in bug reports). We use multimodal GPT-5~\cite{gpt5} to synthesize all available evidences into a structured root-cause diagnosis. We define \texttt{root\_cause} as a concise causal explanation that connects the vulnerability-triggering condition to the vulnerable operation. Each diagnosis includes four parts: (i) \textit{report analysis} integrating CVE description, CWE classification, and visual evidence alongside GitHub issues and project READMEs; (ii) \textit{code-level evidence} correlating commit messages with code diffs; (iii) \textit{contextual insights} synthesizing PR discussion details not present in the initial report; and (iv) \textit{conflict resolution}, which outputs a \texttt{Resolved/Unresolved} label indicating whether the available evidence sufficiently supports the root cause or remains conflicting.
In total, 94 samples flagged as \texttt{Unresolved} underwent human review by two authors with software security expertise, following established practices~\cite{Nong2025APPATCH,Pearce2025Asleep}; the two authors annotated independently (\(\kappa = 0.76\)), and disagreements were resolved by discussion with a third author serving as tiebreaker. Of these, human review resolved 14 cases by reconciling the available evidence, while 80 were excluded due to insufficient evidence for root-cause construction. Furthermore, two authors with software security expertise independently derived root causes for 100 randomly sampled \texttt{Resolved} instances from source code, patches, issue reports, and PR discussions without seeing GPT-5\textquotesingle{}s outputs. Their diagnoses agreed with the summaries on the root cause in all 100 cases. This process yields \textbf{4,998 evidence-based CVE ground-truth reports}, each comprising a \textit{summary} and an \textit{evidence} block. The \textit{summary} serves as the primary evaluation reference, containing a \texttt{root\_cause} narrative, a \texttt{fixing\_pattern} description, and a \texttt{Resolved/Unresolved} label. The \textit{evidence} block records per-source analyses and a \textit{conflict\_resolution} log, providing traceable support for the \textit{summary} conclusions.

\section{Empirical Study}
\label{sec:empirical}

\subsection{Experimental Setup}
\label{sec:setup}

\noindent\textbf{Models.}
We evaluate 10 representative LLMs spanning a wide spectrum of size and capability: six standard models (Qwen2.5-7B, Llama3.1-8B, Llama3.2-3B, GPT-3.5, DeepSeek-V3, and Claude-Sonnet-4) and four reasoning models (QwQ-32B, DeepSeek-R1, Gemini-2.5-Pro, and Grok-4).

\noindent\textbf{Evaluation Datasets.} Evaluating all 4,998 annotated CVEs across ten LLMs would be prohibitively expensive. To ensure reliable reasoning evaluation, we require each test sample to have a root-cause diagnosis strictly verified by human experts. We therefore evaluate on the manually verified subset from \autoref{sec:3.3}, consisting of 100 audited \texttt{Resolved} CVEs and 14 human-resolved \texttt{Unresolved} CVEs. This yielded 114 CVEs across 26 CWEs, resulting in a total of 228 test samples (vulnerable code and corresponding patches). The remaining annotated CVEs are still useful for the training and can be further human-verified to expand the evaluation set.

\noindent\textbf{Detection Stage:} We prompt models to determine whether a given code snippet contains a vulnerability and provide reasoning for their conclusion. We employ two prompting strategies: (1) \textit{Basic Prompt} provides only the code and its context obtained in \autoref{sec:3.2}, and (2) \textit{CWE-Guided Prompt} additionally specifies the CWE type and asks whether a vulnerability of that type exists. For both prompts, models are required to output a structured response containing a binary \texttt{verdict} and a \texttt{root\_cause} field. This structured output aligns model predictions with the ground truth and reduces ambiguity caused by differences in wording or reasoning style.

\begin{table}[t]
    \centering
    \caption{Reasoning evaluation outcome categories.} 
    \resizebox{0.9\linewidth}{!}{
        \begin{tabular}{cccc}
            \toprule[1.2pt]
            
            \textbf{\begin{tabular}[t]{@{}c@{}}Ground Truth\\$(G)$\end{tabular}} & 
            \textbf{\begin{tabular}[t]{@{}c@{}}LLM \\ Prediction $(P)$\end{tabular}} & 
            \textbf{\begin{tabular}[t]{@{}c@{}}Reasoning \\ Correctness $(R)$\end{tabular}} & 
            \textbf{\begin{tabular}[t]{@{}c@{}}Outcome \\ Category\end{tabular}} \\
            \midrule
            1 (Vul.) & 1 (Vul.) & MATCH & TP \\
            1 (Vul.) & 1 (Vul.) & MISMATCH & FN \\
            1 (Vul.) & 0 (Pat.) & -- & FN \\
            \midrule
            0 (Pat.) & 0 (Pat.) & -- & TN \\
            0 (Pat.) & 1 (Vul.) & MISMATCH & FP \\
            0 (Pat.) & 1 (Vul.) & MATCH & TN$^*$ \\
            
            \bottomrule[1.2pt]
        \end{tabular}
    }
    \vspace{0.2em}
    \caption*{\small $^*$Reclassified as TN: reasoning omits the rectified vulnerability or identifies the fixing pattern within the code.}
    \label{tab:metrics}
    \vspace{-1.5em}
\end{table}

\noindent\textbf{Reasoning Evaluation Stage.}
We assess generated reasoning through both automated and manual analysis. For automated evaluation, GPT-5 serves as an LLM judge over all $228 \times 2 \times 10 = 4{,}560$ reasoning outputs (228 test samples, 2 prompting strategies, 10 evaluated models). 
The judge compares the model-generated \texttt{root\_cause} with the ground-truth \texttt{root\_cause} in the corresponding report and outputs \texttt{MATCH} if they are semantically consistent, and \texttt{MISMATCH} otherwise. This criterion assesses agreement with the ground truth on the root cause, rather than the correctness of every intermediate reasoning step. A prediction can still be labeled as \texttt{MATCH} despite differences in reasoning path or wording. Conversely, it is labeled as \texttt{MISMATCH} if it attributes the vulnerability to a different cause, points to the wrong variable/function/path/check, omits an essential causal relation, or provides only a generic explanation not grounded in the code. The full detection and judge prompts, together with representative judge examples illustrating accepted differences in wording/detail and rejected cases with incorrect causal explanations, are provided in the open-access artifact.
To extract qualitative insights and validate this judge, we conducted manual open coding~\cite{khandkar2009open,Wang2025Towards} on 400 stratified-randomly sampled instances (200 \texttt{MATCH}, 200 \texttt{MISMATCH}; equal strata to ensure sufficient statistical power for both verdict types; 95\% confidence, $<$5\% margin of error over $N=4{,}560$), following a three-phase protocol:
(1)~\textit{Initialization}: co-labeling a 40-sample pilot to establish annotation guidelines, excluded from subsequent $\kappa$ computation;
(2)~\textit{Independent annotation}: coding the remaining 360 instances independently (\(\kappa = 0.85\));
(3)~\textit{Resolution}: resolving disagreements via discussion, with a third senior author adjudicating edge cases. Evaluated against this expert consensus, the GPT-5 judge exhibited error rates of only 9\% on \texttt{MISMATCH} and 1.5\% on \texttt{MATCH} (macro-averaged accuracy: 94.75\%), confirming its reliability. 
Finally, this systematic analysis yielded a standardized \textit{codebook} documenting distinct 12 reasoning failure modes, which form the analytical foundation for the qualitative analysis in \autoref{sec:rq3}.

\begin{figure*}[t!]
    \centering
    \includegraphics[width=1\linewidth]{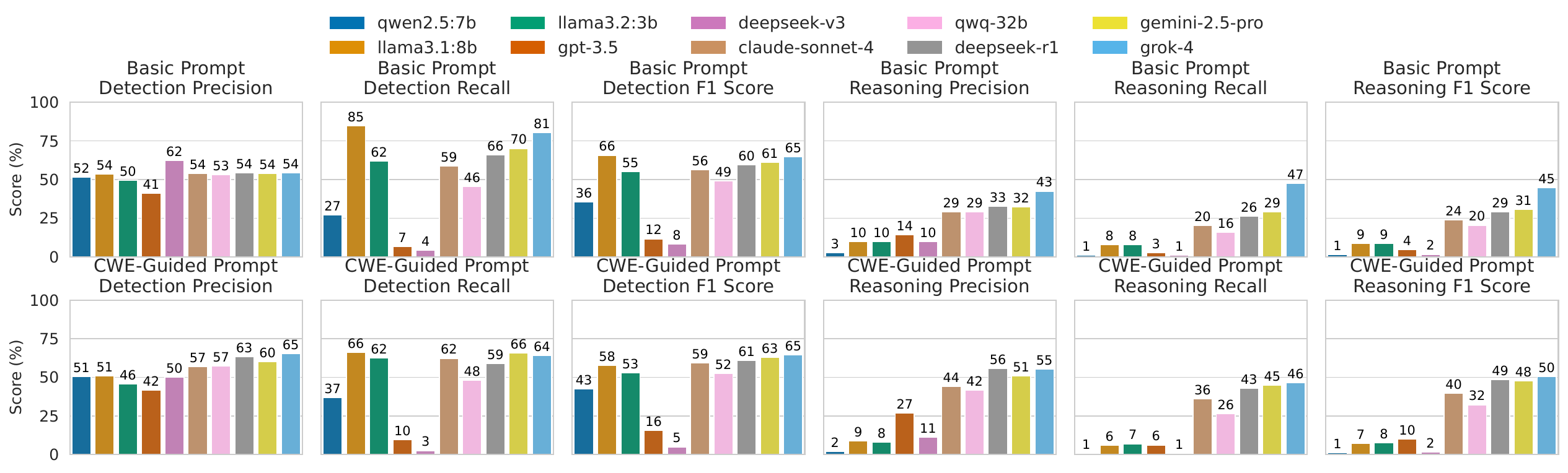}
    \caption{Performance of evaluated LLMs on vulnerability detection and reasoning.
    The 12 subplots compare Detection and Reasoning ($R_{P}$, $R_{R}$, $R_{F1}$) metrics across Basic Prompt and CWE-Guided Prompt strategies.}
    \label{fig:empirical_study_results}
\end{figure*}

\noindent\textbf{Evaluation metrics.}
We define two complementary metric sets to evaluate model performance:

\textbf{\textit{Detection Correctness}:} Standard binary classification metrics (\textbf{accuracy}, \textbf{precision}, \textbf{recall}, \textbf{F1 score}) assess \textbf{binary predictions} without considering reasoning quality, compressing the full analytical process into a single bit.

\textbf{\textit{Reasoning Correctness}:} We introduce reasoning metrics based on the triplet $(G, P, R)$, where $G \in \{0,1\}$ denotes ground truth (0: patched, 1: vulnerable), $P \in \{0,1\}$ denotes prediction, and $R \in \{\texttt{MATCH}, \texttt{MISMATCH}\}$ indicates whether the reasoning aligns with ground truth vulnerability logic. Although $R$ is binary, it distinguishes correct verdicts supported by the target root cause from those attributed to an incorrect cause. Our classification scheme applies asymmetric criteria (\autoref{tab:metrics}):

\begin{itemize}[leftmargin=*]
    \item \textbf{Strict evaluation for vulnerable code ($G=1$):} We require both correct prediction and sound reasoning. Only the triplet (1, 1, $\texttt{MATCH}$) qualifies as a True Positive. Critically, (1, 1, $\texttt{MISMATCH}$) counts as a False Negative despite a correct prediction, since the model fails to identify the actual vulnerability mechanism.
    
    \item \textbf{Lenient evaluation for patched code ($G=0$):} Standard True Negatives include (0, 0, $\text{--}$). However, we reclassify (0, 1, $\texttt{MATCH}$) as True Negative rather than False Positive. This reflects a key insight: when the model identifies alternative concerns or correctly analyzes the efficacy of the patch, this can represent a sophisticated understanding of the code rather than a false alarm regarding the target vulnerability.
\end{itemize}

Using this reasoning-adjusted confusion matrix, we compute \textbf{reasoning accuracy} ($R_{Acc}$), \textbf{reasoning precision} ($R_P$), \textbf{reasoning recall} ($R_R$), and \textbf{reasoning F1 score} ($R_{F1}$).

\begin{figure}[h!]
    \centering
    \includegraphics[width=1\linewidth]{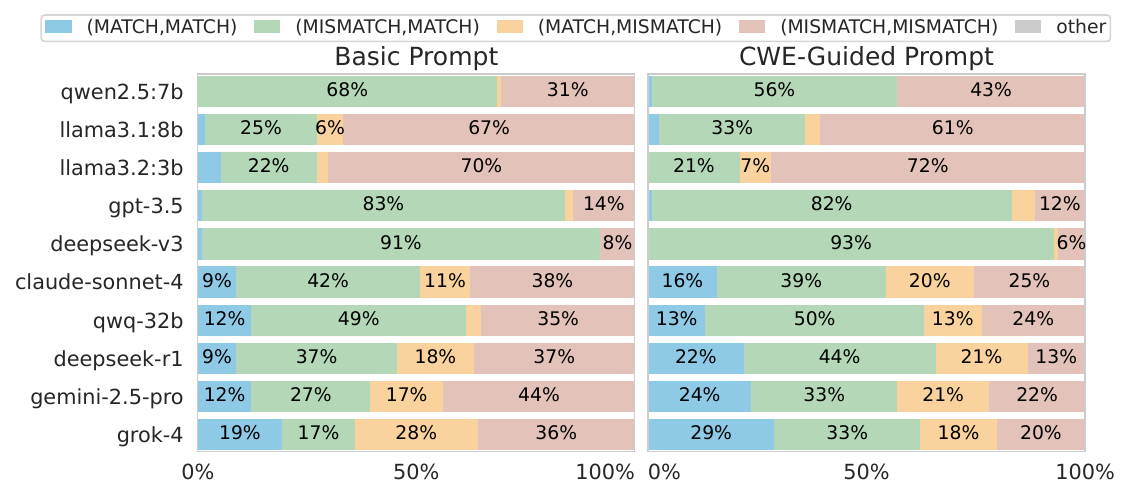}
    \caption{Pair-wise consistency on vulnerable--patched pairs.}
    \label{fig:pair_wise_results}
\end{figure}

\subsection{RQ1: How Well Do Evaluated LLMs Reason About Real-World Vulnerabilities?}
\label{sec:rq1}

\autoref{fig:empirical_study_results} tells a story that binary benchmarks cannot.
Across all 10 evaluated models, \textit{detection F1 scores} are substantially higher than \textit{reasoning F1 scores} (\textit{$R_{F1}$}).
For premier models such as Grok-4, Gemini-2.5-Pro, and DeepSeek-R1, detection F1 scores are around 60--65\%, whereas reasoning F1 scores range from about 29\% to 50\% across the two prompting strategies.
This disparity is even more critical in smaller models.
For instance, under the Basic Prompt setting, Llama-3.1-8b achieves an artificially high \textit{detection recall} of over 80\%, yet its \textit{reasoning recall} (\textit{$R_R$}) is below 10\%.

Among correct binary verdicts, the proportion labeled \texttt{MISMATCH} is 34.6\% (406/1172) under Basic prompting and 26.1\% (319/1222) under CWE-Guided prompting, aggregated across the ten models. Even for Grok-4, this proportion reaches 30.5\% (39/128) under Basic prompting.

\findred{\textbf{The ``Flawed Reasoning, Correct Answer'' Phenomenon.} Across both prompting strategies, 30.3\% (725/2394) of correct binary verdicts from evaluated LLMs are grounded in incorrect reasoning.
Binary accuracy metrics systematically mask this failure, creating a false sense of security in LLM vulnerability analysis.}

\textbf{Results by CWE.} We define the reasoning error rate as the proportion of evaluated outputs labeled \texttt{MISMATCH}. \autoref{tab:cwe_breakdown} shows that reasoning errors are common across the five most frequent CWE categories. Even for seemingly straightforward NULL pointer dereferences (CWE-476), more than half of the outputs fail to identify the correct root cause under either prompting strategy.

\begin{table}[t]
\centering
\small
\setlength{\tabcolsep}{2.5pt}
\newlength{\cwepromptwidth}
\settowidth{\cwepromptwidth}{\textbf{CWE-Guided Prompt}}
\newlength{\cwecountwidth}
\settowidth{\cwecountwidth}{(331/520)}
\caption{Reasoning error rates across the five most frequent CWE categories.}
\label{tab:cwe_breakdown}
\begin{tabular}{cccc}
\toprule
\multirow{2}{*}{\textbf{CWE}} & \multirow{2}{*}{\textbf{CVEs}} & \multicolumn{2}{c}{\textbf{Reasoning Error Rate (\%)}} \\
\cmidrule(lr){3-4}
 & & \makebox[\cwepromptwidth][c]{\textbf{Basic Prompt}} & \makebox[\cwepromptwidth][c]{\textbf{CWE-Guided Prompt}} \\
\midrule
CWE-125 & 26 & 63.65\hspace{3pt}\makebox[\cwecountwidth][c]{(331/520)} & 57.69\hspace{3pt}\makebox[\cwecountwidth][c]{(300/520)} \\
CWE-787 & 16 & 61.88\hspace{3pt}\makebox[\cwecountwidth][c]{(198/320)} & 60.94\hspace{3pt}\makebox[\cwecountwidth][c]{(195/320)} \\
CWE-119 & 12 & 66.67\hspace{3pt}\makebox[\cwecountwidth][c]{(160/240)} & 65.83\hspace{3pt}\makebox[\cwecountwidth][c]{(158/240)} \\
CWE-476 & 10 & 72.50\hspace{3pt}\makebox[\cwecountwidth][c]{(145/200)} & 52.00\hspace{3pt}\makebox[\cwecountwidth][c]{(104/200)} \\
CWE-190 & 6 & 55.83\hspace{3pt}\makebox[\cwecountwidth][c]{(67/120)} & 55.83\hspace{3pt}\makebox[\cwecountwidth][c]{(67/120)} \\
\bottomrule
\end{tabular}
\end{table}

\textbf{Pair-wise consistency: a stricter lens.}
We deepen the analysis by treating each vulnerable snippet and its corresponding patch as a coupled test pair.
A genuinely robust model should simultaneously identify the root cause in the vulnerable version and confirm the safety of the patched version.
\autoref{fig:pair_wise_results} shows that the proportion of (\texttt{MATCH}, \texttt{MATCH}) outcomes is remarkably low across all models.
Even top-tier reasoners such as Grok-4 and DeepSeek-R1 fall below a 30\% pair-wise consistency rate.
This means that when a model correctly identifies a vulnerability, it fails to correctly understand why the patch resolves it more than 70\% of the time. We also observe a positive effect from CWE-Guided prompting: providing the model with the relevant vulnerability class generally improves $R_{F1}$, especially in stronger models.

\find{\textbf{Insight 1: Semantic Instability Across Patch Boundaries.} Less than 30\% of model responses maintain correct reasoning across both the vulnerable and the patched version of the same code.
Even when a model identifies a vulnerability, it typically cannot explain why the fix is effective.}

\subsection{RQ2: Does Reasoning Survive Code Perturbations?}
\label{sec:rq2}

RQ1 establishes that even correct verdicts are frequently backed by flawed reasoning. This raises a follow-up question: \textit{when models reason correctly, is this capability grounded in code semantics, or is it merely a fragile alignment with superficial syntax?} To answer this, we evaluate the models' robustness against semantic-preserving perturbations.
Unlike RQ1, evaluating perturbation robustness requires a controlled environment. We shift from real-world CVEs to the Juliet Test Suite~\cite{meade2018juliet} for two reasons. First, real-world CVEs involve complex interprocedural call chains and build dependencies that make automated perturbation impractical without inadvertently altering vulnerability semantics~\cite{SV-TrustEval-C}; Juliet's self-contained structure permits automated \texttt{g++} compilation checks and straightforward human verification of semantic preservation. Second, Juliet serves as \textit{a conservative lower bound}: if models cannot maintain consistent reasoning across minor surface changes to simple programs, their apparent capability on harder tasks is even less trustworthy.

\begin{table}[t]
\centering
\caption{Reasoning Performance ($R_P$, $R_R$, $R_{F1}$) Across Prompts and Datasets. Arrows in Rephrased Dataset columns indicate performance shifts relative to the Original Dataset (\textcolor{red}{$\downarrow$} for decline, \textcolor{green}{$\uparrow$} for improvement).}
\label{tab:rephrased_results}
\resizebox{\linewidth}{!}{
\begin{tabular}{l|rrr|rrr}
\hline
\multirow{2}{*}{\textbf{Model}} & \multicolumn{3}{c|}{\textbf{Original Dataset}} & \multicolumn{3}{c}{\textbf{Rephrased Dataset}} \\
 & \multicolumn{1}{c}{$R_{P}$} & \multicolumn{1}{c}{$R_{R}$} & \multicolumn{1}{c|}{$R_{F1}$} & \multicolumn{1}{c}{$R_{P}$} & \multicolumn{1}{c}{$R_{R}$} & \multicolumn{1}{c}{$R_{F1}$} \\ \hline \hline
\multicolumn{7}{c}{\textit{\textbf{Panel A: Basic Prompt}}} \\ \hline
qwen2.5:7b & 20.83 & 20.00 & 20.41 & 15.56 \textcolor{red}{$\downarrow$} & 13.46 \textcolor{red}{$\downarrow$} & 14.43 \textcolor{red}{$\downarrow$} \\
llama3.1:8b & 34.21 & 52.00 & 41.27 & 17.74 \textcolor{red}{$\downarrow$} & 21.15 \textcolor{red}{$\downarrow$} & 19.30 \textcolor{red}{$\downarrow$} \\
llama3.2:3b & 25.40 & 32.00 & 28.32 & 9.09 \textcolor{red}{$\downarrow$} & 9.62 \textcolor{red}{$\downarrow$} & 9.35 \textcolor{red}{$\downarrow$} \\
gpt-3.5 & 49.06 & 52.00 & 50.49 & 12.24 \textcolor{red}{$\downarrow$} & 11.54 \textcolor{red}{$\downarrow$} & 11.88 \textcolor{red}{$\downarrow$} \\
deepseek-v3 & 46.77 & 58.00 & 51.79 & 12.50 \textcolor{red}{$\downarrow$} & 11.54 \textcolor{red}{$\downarrow$} & 12.00 \textcolor{red}{$\downarrow$} \\
claude-4-sonnet & 36.67 & 44.00 & 40.00 & 21.88 \textcolor{red}{$\downarrow$} & 26.92 \textcolor{red}{$\downarrow$} & 24.14 \textcolor{red}{$\downarrow$} \\
qwq-32b & 95.35 & 83.67 & 89.13 & 88.57 \textcolor{red}{$\downarrow$} & 60.78 \textcolor{red}{$\downarrow$} & 72.09 \textcolor{red}{$\downarrow$} \\
deepseek-r1 & 78.26 & 72.00 & 75.00 & 77.42 \textcolor{red}{$\downarrow$} & 46.15 \textcolor{red}{$\downarrow$} & 57.83 \textcolor{red}{$\downarrow$} \\
gemini-2.5-pro & 63.33 & 76.00 & 69.09 & 22.64 \textcolor{red}{$\downarrow$} & 23.08 \textcolor{red}{$\downarrow$} & 22.86 \textcolor{red}{$\downarrow$} \\
grok-4 & 64.91 & 74.00 & 69.16 & 22.41 \textcolor{red}{$\downarrow$} & 25.00 \textcolor{red}{$\downarrow$} & 23.64 \textcolor{red}{$\downarrow$} \\
\hline
\multicolumn{7}{c}{\textit{\textbf{Panel B: CWE-Guided Prompt}}} \\ \hline
qwen2.5:7b & 6.45 & 4.00 & 4.94 & 3.03 \textcolor{red}{$\downarrow$} & 1.92 \textcolor{red}{$\downarrow$} & 2.35 \textcolor{red}{$\downarrow$} \\
llama3.1:8b & 35.90 & 56.00 & 43.75 & 20.63 \textcolor{red}{$\downarrow$} & 25.00 \textcolor{red}{$\downarrow$} & 22.61 \textcolor{red}{$\downarrow$} \\
llama3.2:3b & 18.03 & 22.00 & 19.82 & 11.86 \textcolor{red}{$\downarrow$} & 13.46 \textcolor{red}{$\downarrow$} & 12.61 \textcolor{red}{$\downarrow$} \\
gpt-3.5 & 63.16 & 72.00 & 67.29 & 48.00 \textcolor{red}{$\downarrow$} & 46.15 \textcolor{red}{$\downarrow$} & 47.06 \textcolor{red}{$\downarrow$} \\
deepseek-v3 & 59.42 & 82.00 & 68.91 & 44.23 \textcolor{red}{$\downarrow$} & 44.23 \textcolor{red}{$\downarrow$} & 44.23 \textcolor{red}{$\downarrow$} \\
claude-4-sonnet & 50.98 & 52.00 & 51.49 & 45.56 \textcolor{red}{$\downarrow$} & 78.85 \textcolor{green}{$\uparrow$} & 57.75 \textcolor{green}{$\uparrow$} \\
qwq-32b & 91.49 & 86.00 & 88.66 & 78.38 \textcolor{red}{$\downarrow$} & 55.77 \textcolor{red}{$\downarrow$} & 65.17 \textcolor{red}{$\downarrow$} \\
deepseek-r1 & 95.12 & 79.59 & 86.67 & 84.21 \textcolor{red}{$\downarrow$} & 61.54 \textcolor{red}{$\downarrow$} & 71.11 \textcolor{red}{$\downarrow$} \\
gemini-2.5-pro & 96.00 & 96.00 & 96.00 & 88.37 \textcolor{red}{$\downarrow$} & 73.08 \textcolor{red}{$\downarrow$} & 80.00 \textcolor{red}{$\downarrow$} \\
grok-4 & 95.74 & 90.00 & 92.78 & 74.00 \textcolor{red}{$\downarrow$} & 71.15 \textcolor{red}{$\downarrow$} & 72.55 \textcolor{red}{$\downarrow$} \\
\hline
\end{tabular}
}
\end{table}

\begin{table*}[t]
\renewcommand{\arraystretch}{0.9}
\centering
\caption{Taxonomy of vulnerability reasoning failure modes identified through manual open-coding. The errors are classified into four high-level categories based on the stage of the analysis process where the failure occurs.}
\label{tab:error_types}
\resizebox{1\linewidth}{!}{
\begin{tabular}{l l l p{0.7\linewidth}}
\toprule
\textbf{High-Level Category} & \textbf{Abbr.} & \textbf{Error Type} & \textbf{Description} \\
\midrule

\textbf{Analytical Focus Errors} & FE & Analysis Focus Identification Error & The model fails to focus on the code snippet relevant to the vulnerability. \\ \midrule

\multirow{5}{*}{\textbf{\shortstack[l]{Code Structure\\\& Semantics\\Errors}}} 
 & CS1 & Data Flow Misunderstanding & Failure to accurately track data paths across variables, function calls, and object attributes. This includes missing data transformations (e.g., sanitization) or misjudging the data's final state. \\
 & CS2 & Control Flow Misunderstanding & Failure to accurately understand execution paths and conditions, including incorrect judgments on code block reachability, loop termination conditions, or branching logic. \\
 & CS3 & Intra-procedural Semantic Error & Misunderstanding logic within a single function or block (e.g., atomic operations, local calculations) without involving cross-function calls. \\
 & CS4 & Inter-procedural Semantic Error & Misunderstanding semantics across function calls, module boundaries, or interactions with external systems (e.g., threads, files, memory heap). \\ \midrule

\multirow{5}{*}{\textbf{\shortstack[l]{Logic Analysis\\\& Reasoning\\Errors}}} 
 & AR1 & Fallacy of Incomplete Evidence & Creating an argument using only supporting evidence while systematically ignoring key counter-evidence (e.g., existing checks or context) that would refute the conclusion. \\
 & AR2 & Spurious Causality Fallacy & Incorrectly identifying a superficial feature (correlation) of the code as the root cause (causality) of a vulnerability. \\
 & AR3 & Flawed Premise Fallacy & Making incorrect assumptions about the execution environment, trust boundaries, API behaviors, or data sources. \\
 & AR4 & Contradiction Fallacy & Deriving two or more logically conflicting statements within the same reasoning trace. \\ \midrule

\multirow{4}{*}{\textbf{\shortstack[l]{Generative\\Bias}}} 
 & GB1 & Hallucination Bias & Generating information completely inconsistent with the code facts, such as fabricating non-existent variable names, function calls, dependencies. \\
 & GB2 & Over-inference Bias & Making disproportionate extrapolations based on weak clues, constructing extremely complex or practically impossible attack chains. \\
 & GB3 & Redundancy Bias & Generating repetitive, templated, or common-sense information unrelated to the core argument. \\ 

\bottomrule
\end{tabular}
}
\end{table*}

To this end, we randomly stratified-sampled 100 Juliet programs across five representative CWEs (20 per CWE). We first applied \textit{hint elimination}: we removed comments (e.g., \texttt{/* FLAW */}) and label-revealing preprocessor guards (e.g., \texttt{OMITBAD} and \texttt{OMITGOOD}), renamed suggestive functions, and replaced print strings that revealed vulnerability labels. We then applied 26 semantic-preserving transformations (e.g., variable renaming and dead code insertion) following CodeMorph~\cite{Rao2025CODEMORPH}, which maximizes surface divergence while preserving semantics, yielding 100 perturbed instances (see full details in the open-access artifact). 
Through \textit{vulnerability-centric locking}, two authors annotated the root-cause lines to keep them invariant during perturbation. Every perturbed instance then passed a dual-gate validation comprising successful compilation and human review, with two authors independently verifying semantic preservation (\(\kappa = 0.88\)); the 5 instances that failed were revised and re-validated, with persistent disagreements adjudicated by a third author until consensus was reached. We use the same detection and reasoning evaluation prompts as in RQ1; the reasoning ground truth is reconstructed for Juliet by combining its official comments with the annotated root-cause lines. Since the core vulnerability logic remains unchanged by construction, any performance degradation directly indicates over-reliance on vulnerability-irrelevant syntactic features.

\autoref{tab:rephrased_results} documents the results.
The degradation is substantial and universal.
Under Basic prompting, $R_{F1}$ drops by an average of 26.7\% across all evaluated models.
Notably, under CWE-Guided prompting, Claude-Sonnet-4 exhibits an unexpected $R_{F1}$ increase (from 51.49\% to 57.75\%). 
Manual inspection reveals that the model explicitly recognizes the injected transformations, suggesting the presence of defensive anti-perturbation training.
Although CWE-Guided prompting attenuates this degradation by reducing the average $R_{F1}$ drop to 14.5\%, the remaining loss in reasoning performance under semantic-preserving changes is still substantial.

\find{\textbf{Insight 2: Reasoning Shows Fragility Under Perturbation.} 
Semantic-preserving transformations reduce $R_{F1}$ by 26.7\% under Basic prompting and by 14.5\% under CWE-Guided prompting. This suggests that current LLM vulnerability reasoning remains sensitive to surface-level code changes.}

\subsection{RQ3: What Exactly Goes Wrong in Reasoning?}
\label{sec:rq3}

\begin{figure}[htbp]
    \centering
    \includegraphics[width=1\linewidth]{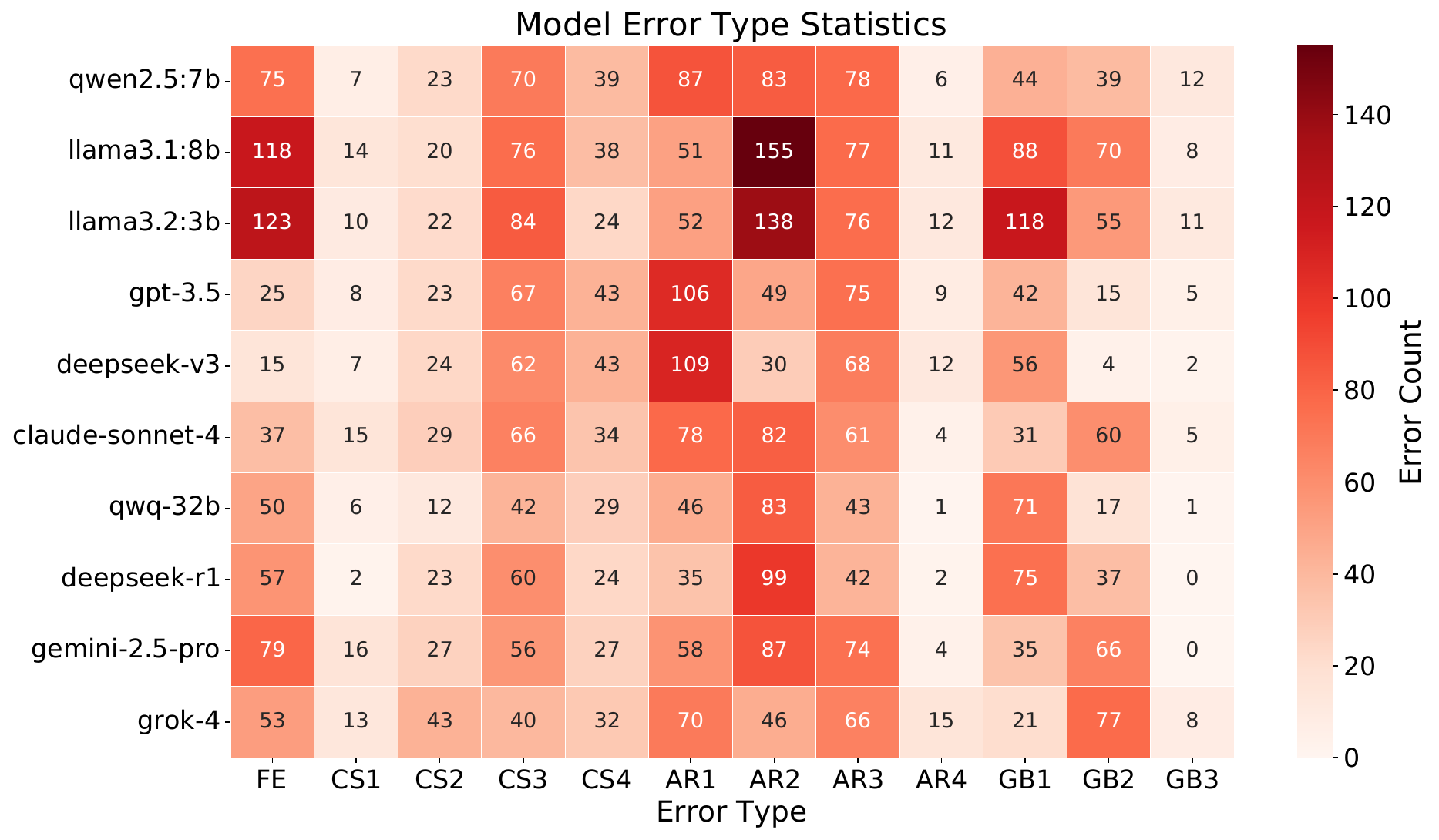}
    \caption{Distribution of error types across 10 LLMs.}
    \label{fig:model_error_heatmap}
\end{figure}

We have established that LLM vulnerability reasoning fails at an alarming rate and that failures are amplified by surface-level code changes. The next question is: \textit{how does it fail?} Drawing on open coding of flawed reasoning instances in our 400-sample manual validation set (\autoref{sec:setup}), we constructed a codebook of \textbf{12 failure modes} organized into four dimensions: analytical focus, code structure and semantics, logical analysis, and generative bias. Notably, the taxonomy excludes the 14 rare instances of correct reasoning with incorrect labels (the \texttt{TN$^*$} scenario in \autoref{tab:metrics}). Although these patched instances predict a vulnerable label, they demonstrate genuine code comprehension (e.g., accurately analyzing patch efficacy or identifying alternative concerns) rather than reasoning failure. \autoref{tab:error_types} presents the full taxonomy, while \autoref{fig:model_error_heatmap} shows their distribution across models. Representative examples for these error types are provided in the artifact.

To analyze the distribution of 12 failure modes at scale, we use GPT-5 to evaluate the 2,846 instances of flawed reasoning (\texttt{MISMATCH}) identified across all models ($N=4{,}560$). Before doing so, we first evaluate the reliability of using GPT-5 for multi-label error classification on the manually annotated failure instances. The evaluation yields a Jaccard Index of 0.73. Furthermore, to account for chance agreement, we calculate the macro-averaged Cohen's $\kappa$, which reaches a score of 0.72. This confirms a substantial agreement between the automated classifier and human experts. With this validation, we use GPT-5 along with few-shot prompting to categorize the failures across all evaluated models according to our established codebook.
As shown in \autoref{fig:model_error_heatmap}, \textbf{CS and AR errors dominate the heatmap for every evaluated model}. For instance, AR1 and CS3 consistently exhibit high frequencies, with AR1 reaching 109 in DeepSeek-V3 and CS3 hitting 66 in Claude-Sonnet-4. This means failures in taint tracking, constraint solving, and evidentiary completeness are pervasive. 
Interestingly, generative bias errors are less frequent than CS and AR errors, and often decrease in advanced models (e.g., 21 for GB1 in Grok-4 and 0 for GB3 in Gemini-2.5-Pro). This indicates LLMs largely remain faithful to the original code context. Therefore, a key challenge to mitigate in LLM vulnerability analysis is the difficulty of maintaining logical consistency during complex semantic code comprehension.

A further observation from open coding is that reasoning errors often co-occur rather than appearing in isolation. For example, a control-flow misinterpretation (CS2) may supply an incorrect premise for later analysis, which then leads to an unsupported causal claim (AR1). This suggests that many failures stem from broken logical consistency across dependent reasoning steps, rather than from a single isolated mistake.

\findblue{\textbf{Transition: From Diagnosis to Treatment.} RQ1--RQ3 tell a consistent story: LLMs frequently produce unreliable vulnerability reasoning (RQ1), correct reasoning often relies on surface features rather than genuine code comprehension (RQ2), and reasoning failures are dominated by code comprehension and logical reasoning errors (RQ3). Together, they expose the difficulty of maintaining evidence-grounded logical consistency under complex code semantics.}

\section{Enhancement Method: \tool{}}
\label{sec:dagvul}

The empirical evidence points to a key challenge in evidence-grounded logical consistency: models struggle to connect code facts, semantic constraints, and final vulnerability conclusions. CoT leaves these dependencies implicit, making unsupported reasoning steps hard to detect or optimize during training.
We propose \tool{}, a framework that addresses this limitation by reformulating vulnerability reasoning as a directed acyclic graph (DAG) generation task and aligning the generation process with program-intrinsic logic through Reinforcement Learning with Verifiable Rewards (RLVR).

\subsection{Modeling the \tool{}}
\label{sec:modeling}

\begin{figure*}[t]
    \centering
    \includegraphics[width=0.9\linewidth]{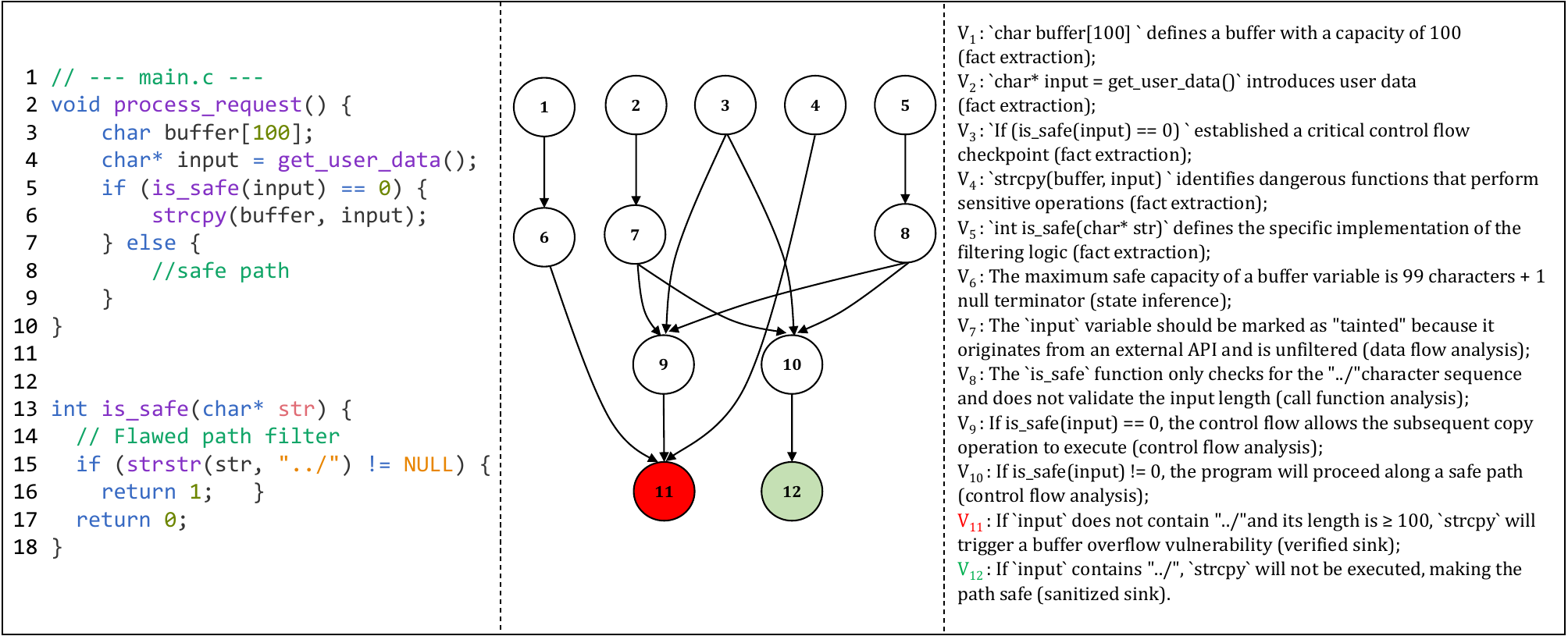}
    \caption{Illustrative example of the \tool{} reasoning graph.
    Left: a simplified C code with a flawed check.
    Center: the inferred DAG topology.
    Right: semantic descriptions of nodes $v_1$--$v_{12}$, generated following the \tool{} protocol.}
    \label{fig:dag_example_full}
\end{figure*}

We formalize the vulnerability reasoning process as a DAG $G(x_{in}) = (V(x_{in}), E(x_{in}))$, where $x_{in}$ is the input code, $V$ is the set of reasoning nodes, and $E$ is the set of directed dependency edges between them. This builds on graph-based reasoning formalizations in prior work~\cite{zhang2025dagmath}, specialized here for the demands of security analysis.

\textbf{Node types.}
Nodes in $V(x_{in})$ play three distinct roles.
\textit{Source nodes} ($V_{in}$) encode vulnerability-relevant ground facts extracted directly from source code (buffer capacities, API definitions, entry points) with no derived inference permitted; they must contain only literal code facts.
\textit{Intermediate nodes} ($V_{inter}$) represent logical inferences built from preceding steps, encompassing taint tracking, pointer analysis, and constraint solving.
\textit{Sink nodes} ($V_{out}$) carry the final audit verdict and are split into two subtypes: a \textit{verified\_sink} ($v_{11}$ in \autoref{fig:dag_example_full}) concluding a confirmed vulnerability with a full exploit path, and a \textit{sanitized\_sink} ($v_{12}$) formally asserting code safety by proving the efficacy of existing sanitization or guards.
The dual terminal structure reflects a key insight from modern security auditing: proving safety is as critical as identifying flaws.

\textbf{Edge constraints.}
Two principles govern edge validity.
The \textit{program analysis principle} requires each edge to explicitly name the analysis primitive used (e.g., taint propagation, data flow analysis, constraint solving).
The \textit{citation rule} mandates that every reasoning justification reference its antecedents by the exact tag ``Step X'', making every inferential leap auditable.

\textbf{Admissibility constraint.}
The DAG structure is operationally enforced during generation.
Let $v_{1:t-1}$ be the nodes generated from step 1 to $t-1$, and $pa(v)$ denote the parent nodes of $v$.
The admissible set for step $t$ is:
\begin{equation}
\begin{aligned}
\mathcal{V}(v_{1:t-1} \,|\, x_{in}) :=
\{\, v \in V(x_{in}) :\;& pa(v) \subseteq \{v_{1:t-1}\}, \\
& v \notin \{v_{1:t-1}\} \,\}
\end{aligned}
\end{equation}
This rule enforces a strict logical order; a node can only be visited once all its parent dependencies have been established in the preceding context. As shown in our example in \autoref{fig:dag_example_full}, after the model has collected $\{v_1, v_2, v_3, v_4, v_5, v_6, v_7\}$, the only logically admissible next step is the state inference node $v_8$. Nodes $v_9$ through $v_{12}$ remain inaccessible until the dependency $v_8$ is resolved.

\textbf{Evaluation metrics within the DAG.}
First, \textit{Logical Closeness} dictates that every path originating from the source nodes ($V_{in}$) terminates at a defined terminal sink ($V_{out}$). This structural constraint ensures that no intermediate assertions are left ``dangling'' and that the final verdict is derived from the preceding reasoning steps. 
Second, \textit{Reasoning Correctness} assesses the alignment between the generated sink nodes and the ground truth.
Specifically, the presence of any \textit{verified\_sink} indicates the discovery of a flaw, whereas a reasoning graph whose sink nodes are all \textit{sanitized\_sinks} confirms code safety.

\subsection{Gold Standard Dataset}
\label{sec:gold_standard_dataset}

Generating complex DAG structures directly from source code is inherently challenging~\cite{zhang2025dagmath}. We therefore use CoT trajectories as a structural scaffold for our synthesis pipeline, applied to the 4,998 samples collected in \autoref{sec:benchmark}.

\textbf{CoT Trajectory Generation.} We prompt DeepSeek-R1 to generate CoT trajectories (\autoref{fig:dag_example_full} shows an example) anchored to our evidence-based ground truth. To ensure that the model infers the root cause from evidence in the code rather than merely copying the answer, we apply strict filtering: we reject responses containing explicit ground-truth terminology or exceeding a 0.8 semantic similarity threshold (measured via Qwen3-Embedding-0.6B, chosen for its SOTA performance on MTEB~\cite{mteb_leaderboard}). This threshold was manually calibrated by security experts on a pilot subset. We allow three resampling attempts before discarding a sample.

\textbf{DAG Synthesis.} Given the source code, its interprocedural context, and the filtered CoT trajectory, we prompt DeepSeek-R1 with carefully designed prompts to synthesize the DAG in three stages: (1) \textbf{Node Identification}, (2) \textbf{Dependency Assignment}, and (3) \textbf{Edge Generation} (detailed prompts are available in our artifact). Critically, the structural correctness of the resulting graphs is not contingent on the LLM's reliability: all graph constraints are enforced by deterministic rule-based algorithms operating on explicitly defined node types and edge relations. Specifically, we prune dangling nodes ($v \in V_{inter} \cup V_{in}$) that form vulnerability-irrelevant reasoning paths not reaching any sink nodes. We then discard samples where an intermediate node ($v \in V_{inter}$) leading to a \textit{verified\_sink} lacks parent dependencies, as such reasoning lack support from evidence-grounded code facts.
Two authors independently reviewed 200 randomly sampled DAGs to assess whether the node semantics and dependency edges faithfully reflected the underlying vulnerability logic.
The review found no invalid samples, providing additional evidence for the reliability of the synthesis pipeline.

The final Gold Standard Dataset comprises \textbf{4,041 vulnerability samples} and \textbf{4,116 patch samples} (full topological statistics are provided in our artifact). Notably, the high-quality CoT trajectories generated during this process are preserved as a training corpus for SFT and ORPO baselines.

\definecolor{BestColor}{RGB}{200, 200, 240}
\definecolor{SecondBestColor}{RGB}{230, 230, 250}

\begin{table*}[t]
  \centering
  \renewcommand{\arraystretch}{0.9}
  \setlength{\fboxsep}{1pt}
  \caption{Performance of Qwen3, Llama3.1-Nemotron, and Claude-Sonnet-4.5 across different scales. We evaluate Reasoning metrics under Basic and CWE-Guided prompts. (\colorbox{BestColor}{best}, \colorbox{SecondBestColor}{second-best} $R_{F1}$ within each model group).}
  \label{tab:enhancement_results}
  \vspace{0.2cm}
  
  \resizebox{0.8\linewidth}{!}{%
  \begin{tabular}{l|l|l|cccc|cccc}
    \toprule[1.2pt]
    \multirow{2}{*}{\textbf{Base Model}} & \multirow{2}{*}{\textbf{Inference Strategy}} & \multirow{2}{*}{\textbf{Optimization Method}} & \multicolumn{4}{c|}{\textbf{Basic Prompt}} & \multicolumn{4}{c}{\textbf{CWE-Guided Prompt}} \\
    \cmidrule(lr){4-11}
    & & & $R_{Acc}$ & $R_{P}$ & $R_{R}$ & \textbf{$R_{F1}$}$^\dagger$ & $R_{Acc}$ & $R_{P}$ & $R_{R}$ & \textbf{$R_{F1}$}$^\dagger$ \\
    \midrule

    \multirow{10}{*}{Qwen3-8B} 
    & Simple$^*$ & None & 53.51 & 53.45 & 54.39 & 53.91 & 69.91 & 92.31 & 42.86 & 58.54 \\
    & CoT    & None & 59.65 & 57.14 & 77.19 & \colorbox{SecondBestColor}{65.67} & 61.06 & 59.09 & 69.64 & 63.93 \\
    & CoT    & RAG      & 57.02 & 55.71 & 68.42 & 61.42 & 64.60 & 62.32 & 75.44 & 68.25 \\
    & CoT    & SFT      & 52.63 & 51.58 & 85.96 & 64.47 & 64.04 & 61.11 & 77.19 & 68.22 \\
    & CoT    & ORPO     & 50.00 & 50.00 & 82.46 & 62.25 & 64.91 & 61.97 & 77.19 & 68.75 \\
    & CoT    & GRPO     & 54.39 & 53.73 & 63.16 & 58.06 & 69.30 & 72.00 & 63.16 & 67.29 \\
    & DAG    & None     & 62.83 & 65.31 & 56.14 & 60.38 & 60.18 & 57.32 & 82.46 & 67.63 \\
    & DAG    & SFT (Cold Start)     & 67.54 & 72.73 & 56.14 & 63.37 & 73.68 & 81.40 & 61.40 & 70.00 \\
    & DAG    & \textbf{\tool{} (Our Method)} & 76.11 & 91.67 & 57.89 & \colorbox{BestColor}{70.97} & 76.32 & 83.78 & 72.09 & \colorbox{BestColor}{77.50} \\
    & DAG    & w/o Logic Reward & 68.42 & 73.33 & 57.89 & 64.71 & 77.19 & 81.63 & 70.18 & \colorbox{SecondBestColor}{75.47} \\
    & DAG    & w/o Reasoning Reward & 63.01 & 80.00 & 47.62 & 59.70 & 71.43 & 80.00 & 65.12 & 71.79 \\
    
    \midrule

    \multirow{4}{*}{Qwen3-8B-Reasoning}
    & Simple$^*$ & None & 64.91 & 66.67 & 59.65 & 62.96 & 71.05 & 76.09 & 61.40 & 67.96 \\
    & CoT    & None & 65.79 & 64.52 & 70.18 & \colorbox{BestColor}{67.23} & 71.05 & 74.00 & 64.91 & \colorbox{BestColor}{69.16} \\
    & CoT    & RAG      & 57.89 & 55.84 & 75.44 & 64.18 & 70.18 & 71.70 & 66.67 & \colorbox{SecondBestColor}{69.09} \\
    & DAG    & None & 66.07 & 67.27 & 64.91 & \colorbox{SecondBestColor}{66.07} & 68.14 & 67.24 & 69.64 & 68.42 \\
    
    \midrule
    
    \multirow{10}{*}{Llama3.1-Nemo-8B}
    & Simple$^*$ & None & 33.33 & 35.82 & 42.11 & 38.71 & 50.00 & 50.00 & 57.89 & 53.66 \\
    & CoT    & None & 48.67 & 47.50 & 33.93 & 39.58 & 49.12 & 49.21 & 54.39 & 51.67 \\
    & CoT    & RAG      & 56.14 & 55.93 & 57.89 & 56.90 & 50.00 & 50.00 & 57.89 & 53.66 \\
    & CoT    & SFT      & 54.39 & 54.72 & 50.88 & 52.73 & 58.77 & 56.58 & 75.44 & 64.66 \\
    & CoT    & ORPO     & 52.63 & 52.94 & 47.37 & 50.00 & 59.65 & 57.33 & 75.44 & 65.15 \\
    & CoT    & GRPO     & 62.28 & 66.67 & 49.12 & 56.57 & 65.79 & 67.31 & 61.40 & 64.22 \\
    & DAG    & None     & 53.27 & 53.85 & 26.92 & 35.90 & 41.51 & 44.00 & 62.26 & 51.56 \\
    & DAG    & SFT (Cold Start)     & 60.53 & 63.04 & 50.88 & 56.31 & 69.30 & 70.37 & 66.67 & 68.47 \\
    & DAG    & \textbf{\tool{} (Our Method)} & 67.54 & 76.32 & 50.88 & \colorbox{BestColor}{61.05} & 79.82 & 82.69 & 75.44 & \colorbox{BestColor}{78.90} \\
    & DAG    & w/o Logic Reward & 65.49 & 70.73 & 51.79 & \colorbox{SecondBestColor}{59.79} & 74.34 & 78.00 & 68.42 & \colorbox{SecondBestColor}{72.90} \\
    & DAG    & w/o Reasoning Reward & 65.79 & 73.68 & 49.12 & 58.95 & 73.45 & 77.08 & 66.07 & 71.15 \\

    \midrule

    \multirow{4}{*}{Llama3.1-Nemo-8B-Reasoning}
    & Simple$^*$ & None & 34.21 & 37.14 & 45.61 & 40.94 & 49.12 & 49.25 & 57.89 & 53.23 \\
    & CoT    & None & 46.49 & 45.45 & 35.09 & 39.60 & 49.56 & 49.21 & 55.36 & 52.10 \\
    & CoT    & RAG      & 56.14 & 55.74 & 59.65 & \colorbox{BestColor}{57.63} & 50.00 & 50.00 & 59.65 & \colorbox{SecondBestColor}{54.40} \\
    & DAG    & None & 51.40 & 51.16 & 41.51 & \colorbox{SecondBestColor}{45.83} & 48.04 & 48.53 & 64.71 & \colorbox{BestColor}{55.46} \\

    \midrule

    \multirow{4}{*}{Qwen3-30B-Reasoning}
    & Simple$^*$ & None & 57.94 & 57.78 & 50.00 & 53.61 & 71.03 & 78.05 & 59.26 & 67.37 \\
    & CoT    & None & 54.08 & 54.29 & 39.58 & 45.78 & 72.63 & 84.38 & 56.25 & \colorbox{SecondBestColor}{67.50} \\
    & CoT    & RAG      & 72.73 & 87.88 & 55.77 & \colorbox{SecondBestColor}{68.24} & 76.92 & 87.18 & 64.15 & \colorbox{BestColor}{73.91} \\
    & DAG    & None & 75.00 & 94.44 & 54.84 & \colorbox{BestColor}{69.39} & 71.23 & 76.92 & 57.14 & 65.57 \\

    \midrule
    
    \multirow{4}{*}{Claude-Sonnet-4.5}
    & Simple$^*$ & None & 69.30 & 68.97 & 70.18 & 69.57 & 70.18 & 74.47 & 61.40 & 67.31 \\
    & CoT    & None & 70.18 & 69.49 & 71.93 & 70.69 & 69.03 & 66.15 & 76.79 & \colorbox{SecondBestColor}{71.07} \\
    & CoT    & RAG      & 74.56 & 78.00 & 68.42 & \colorbox{SecondBestColor}{72.90} & 76.32 & 76.79 & 75.44 & \colorbox{BestColor}{76.11} \\
    & DAG    & None & 73.68 & 72.13 & 77.19 & \colorbox{BestColor}{74.58} & 67.54 & 66.67 & 70.18 & 68.38 \\

    \bottomrule[1.2pt]
  \end{tabular}%
  }
  \captionsetup{justification=raggedright, singlelinecheck=false}
  \vspace{0.2em}
  \caption*{\small $^*$Simple prompting: directly ask the model to identify vulnerabilities, e.g., ``identify if the code is vulnerable and explain the reasoning''. $^\dagger$$R_{F1}$ is the most important metric as it balances precision and recall in vulnerability reasoning.}
\end{table*}

\subsection{Two-Stage Training: Cold Start and RLVR}
\label{sec:training}

\noindent\textbf{Stage 1: Cold-start SFT.}
We initialize training by performing supervised fine-tuning on the complete Gold Standard Dataset ($N = 8,157$) for 3 epochs.
This stage provides the model with a strong prior over the DAG format and the underlying program analysis semantics, establishing a stable starting point for the subsequent reinforcement learning phase.

\noindent\textbf{Stage 2: RLVR with GRPO.}
Building on the SFT model, we apply GRPO~\cite{shao2024deepseekmath} with a rollout size of 16, following standard practice in recent reasoning-oriented RL training~\cite{shao2024deepseekmath,Guo_2025}, to align the reasoning process with structural integrity and causal correctness.
The hybrid reward function is:
\begin{equation}
R_{total} = \omega_1 \cdot \delta_{close}(\mathcal{G}_{gen}) + \omega_2 \cdot \delta_{final}(\mathcal{G}_{gen})
\end{equation}

where $\delta_{close}$ is a rule-based indicator verifying the logical closeness of the generated graph structure, and $\delta_{final}$ is a model-based indicator evaluated by DeepSeek-R1 judging whether the generated sink node aligns with ground truth.
To maintain computational feasibility, the RL subset is filtered to samples
within a 4,096-token context window, yielding 2,178 high-quality training
instances; RLVR is trained for 2 epochs. All training was conducted on a server equipped with 8 NVIDIA A800 GPUs, each with 40GB of memory.

\subsection{RQ4: Does \tool{} Mitigate Reasoning Failures?}
\label{sec:rq4}

We evaluate \tool{} on a deduplicated subset of SVEN~\cite{SVEN}, a widely-adopted real-world benchmark~\cite{lekssays2025llmxcpgcontextawarevulnerabilitydetection,wen2025boostingvulnerabilitydetectionllms,10.1109/ICSE55347.2025.00038} containing 417 samples across 9 prevalent CWEs covered by our training data. We remove 303 samples sharing commit URLs with our training data, leaving 114 code samples for evaluation, and construct the reasoning ground truth following \autoref{sec:3.3}.
Non-closed DAGs with parseable sink explanations are evaluated using the same reasoning judge criteria as closed DAGs. Outputs without valid sink explanations are excluded from the corresponding metric denominators. For Qwen3-8B with \tool{} under Basic prompting, one of the 114 outputs is excluded for this reason.
Baselines include RAG (retrieving top-3 similar code with root causes from our training corpus, following Vul-RAG~\cite{Vul-RAG}), SFT (following VulnLLM-R~\cite{VulnLLM-R}), and ORPO (rejected responses generated via incorrect labels, following R2Vul~\cite{R2Vul}). These methods are applied to Qwen3-8B and Llama3.1-Nemo-8B. We also include a CoT-based GRPO baseline for both models, using the same training hyperparameters and reasoning reward as \tool{}. We also include Qwen3-30B-Reasoning and Claude-Sonnet-4.5 as large-parameter reference. We do not apply the training methods to the reasoning models to avoid disrupting their intrinsic patterns~\cite{Guo_2025}.

As shown in \autoref{tab:enhancement_results}, \textbf{\tool{} consistently achieves the highest reasoning performance} across almost every tested configuration. On Qwen3-8B with Basic prompting, \tool{} reaches $R_{F1} = 70.97\%$, outperforming Simple inference (53.91\%), SFT (64.47\%), and ORPO (62.25\%) by substantial margins. Compared to all baselines within the same model group, \tool{} achieves an average reduction of 24.79\% in the \textit{false alarm rate}. On Qwen3-8B under Basic prompting, \tool{} achieves reasoning precision of 91.67\%, but reasoning recall drops from SFT's 85.96\% to 57.89\%, reflecting a precision--recall trade-off. Meanwhile, RAG exhibits notable instability: its $R_{F1}$ on Qwen3-8B-Reasoning (64.18\%) falls below the CoT baseline (67.23\%), suggesting that retrieved context often introduces distracting noise that degrades reasoning.

On Qwen3-8B under Basic prompting, \tool{} improves $R_{F1}$ from 63.37\% to 70.97\%. The ablation results suggest that the logic reward mainly improves precision, increasing $R_P$ from 72.73\% to 80.00\% when retained alone, while its lower $R_R$ indicates that reasoning reward is necessary for recovering more complete vulnerability evidence. \tool{} also consistently outperforms CoT-based GRPO, suggesting that the gains cannot be attributed to GRPO alone.

\textbf{\tool{} is competitive with larger models.} Under Basic prompting, Qwen3-30B-Reasoning achieves its best $R_{F1}$ of 69.39\%, while DAGVUL on an 8B base model reaches 70.97\%. Under CWE-Guided prompting, \tool{} trained on Llama3.1-Nemo-8B reaches 78.90\%, comparable to Claude-Sonnet-4.5's best result of 76.11\% in this setting.

We also provide case studies in the artifact to qualitatively illustrate how \tool{} mitigates typical reasoning failures, such as missing links between guards and vulnerable operations and drawing conclusions from incomplete code evidence.

\find{\textbf{Insight 3: Structured Reasoning with Task-Specific Training.} Across all baselines, \tool{} yields an average 13.3\% improvement in $R_{F1}$. The results suggest that structured reasoning and evidence-based training can help models connect code facts with vulnerability conclusions.}

\section{Threats to Validity}

\noindent\textbf{Judge reliability.} To ensure scalability across the 4,560 instances, we utilize GPT-5 as an automated judge. To mitigate evaluation bias, two authors validated 400 stratified-randomly sampled instances. The GPT-5 judge exhibited error rates of 9.0\% on \texttt{MISMATCH} and 1.5\% on \texttt{MATCH} verdicts, confirming its reliability. Nonetheless, the LLM-as-a-judge paradigm may still introduce subtle subjective biases relative to expert review.

\noindent\textbf{Ground truth reliability.} Our 4,998 evidence-based CVE ground-truth reports are constructed by GPT-5 through reasoning over multi-source evidence and may not be fully accurate, particularly for \texttt{Resolved} instances that were not exhaustively reviewed by humans. Although such evidence-driven reasoning is unlikely to be entirely free of annotation noise, our manual audit (detailed in \autoref{sec:3.3}) found that all 100 randomly sampled \texttt{Resolved} instances had root-cause diagnoses supported by the available evidence, suggesting that residual noise is unlikely to materially affect our conclusions.

\noindent\textbf{Perturbation generalizability.} In RQ2, we evaluated model robustness against semantic-preserving perturbations using the Juliet Test Suite. We acknowledge that synthetic datasets lack the interprocedural dependencies of real-world software, and the perturbation results may not fully generalize to complex real-world CVEs. Nonetheless, the controlled setting enables clean isolation of reasoning failures attributable to surface-level features, independent of confounding factors present in real-world code.

\section{Discussion}
\label{sec:discussion}

\noindent \textbf{Logical closeness rate.} We measured the percentage of instances achieving logical closeness (as defined in \autoref{sec:modeling}). This metric exhibits significant improvement through \tool{}. Specifically, Qwen3-8B increased from 32.90\% to 73.25\%, while Llama-Nemo-8B rose from 44.74\% to 67.99\%. These results demonstrate that structural constraints effectively guide models to complete end-to-end reasoning chains.

\noindent\textbf{Mitigating reward hacking.}
The asymmetric difficulty between vulnerable and patched code creates a potential reward hacking pathway: a model could maximize $R_{total}$ by aggressively classifying everything as safe, trivially producing sanitized sinks without genuine analysis.
We mitigate this by assigning positive rewards for correctly identifying verified sinks and requiring reasoning veracity verification via the $\delta_{final}$ signal.
Post-training analysis confirms that the models do not degenerate toward a safe-class bias.

\noindent\textbf{Agentic vulnerability reasoning.}
We also performed an exploratory evaluation of agentic systems under the Basic prompt setting. By autonomously retrieving repository context, Claude Code~\cite{claudecode} and Codex~\cite{codex} achieve $R_{F1}$ scores of 78.48\% and 81.40\%, respectively. However, evaluating agent reasoning requires considering additional factors such as tool invocation and context selection, while extending our framework to agentic RL would require larger resources due to long-horizon tool interactions. We therefore leave evaluation and enhancement of agentic vulnerability reasoning to future work.

\section{Conclusion}

This paper argues that binary verdict accuracy is an insufficient measure of LLM-based vulnerability analysis. Through new reasoning-level datasets and large-scale empirical study, we expose 12 reasoning failure modes invisible to existing evaluations. Our proposed \tool{} addresses these failures by modeling vulnerability reasoning as a DAG generation task with RLVR alignment, achieving an average 13.3\% improvement in reasoning F1-score. These results support combining structured causal reasoning with training on curated, task-specific data to improve vulnerability analysis.

\section*{Data Availability}
We have made the experimental data publicly accessible at \url{https://github.com/security-pride/DAGVul}.

\section*{Acknowledgments}
This work was supported in part by the National Natural Science Foundation of China (Grants No. 62502168, 62572209), the Hubei Provincial Key Research and Development Program (Grant No. 2025BAB057), and the International Science and Technology Cooperation Program of Hubei Province (Grant No. 2026EHA039).

\bibliographystyle{IEEEtran}
\bibliography{main}

\end{document}